\documentstyle[epsfig]{elsart}

\begin{document}

\begin{frontmatter}
\title{A novel current mode operating  beam counter\\
       based on not preamplified HPDs}
\author[TOKYO]{M.C. Fujiwara},
\author[CERN]{M. Marchesotti\thanksref{corr_author}}
\address[TOKYO]{University of Tokyo,7-3-1, Hongo, Bunkyo-ku, Tokyo 113, Japan}
\address[CERN]{CERN, PS Division, 1211 Geneva 23, Switzerland}
\thanks[corr_author]{Corresponding author. Tel.:+41-22-7674834,
fax: +41-22-7678955 \\e-mail: Marco.Marchesotti@cern.ch}

\begin{abstract}
A novel system to monitor the intensity and the stability of a bunched beam
of $\simeq 1.2\times 10^{7}$ antiprotons ($\bar{p}$s) with a length of
$\simeq$ 250 ns (FWHM) and to measure its trapping efficiency in a Penning
trap is described. This system operates parasitically detecting the pions
from the annihilation of part of the beam in a degrader.\\
Six plastic scintillators have been coupled from one side to six proximity
focused HPDs without preamplifiers and operating in current mode. This device
works in the stray field of the ATHENA magnet with no loss of efficiency; the
gain can be varied from zero to a few thousands with a precision better than
$0.1\%$ and the dynamic range is larger than 8 orders of magnitude. Linearity
and stability have been measured up to charge responses of 100 nC,
corresponding to the beam completely dumped. The beam counter has been
calibrated in two different and independent ways giving consistent results.
\end{abstract}

\end{frontmatter}

%%%%

\section{Introduction}
The goal of the ATHENA (ApparaTus for High precision Experiments on Neutral
Antimatter) \cite{ATHE1} experiment at CERN, is to produce and store low
energy anti-hydrogen atoms ($\bar{H}$s). The 1S-2S transition can be excited
resonantly and compared with that of the hydrogen atom ($H$) to a high
precision \cite{Mike_Charl}. Any difference may be due either to CPT violation
or to an anomalous red shift originated by a different gravitational
interaction of antimatter \cite{GRAV}.\\
A bunched beam of $\simeq 1.2 \times 10^{7}\,\bar{p}$s with a length of
$\simeq$ 250 ns (FWHM) and a momentum of 100 MeV/c delivered by the Antiproton
Decelerator (AD) \cite{AD1} is degraded before entering a multi-ring Penning
trap. Only $\simeq 0.1\%$ of the beam is trapped, cooled to sub-eV level by
electron cooling \cite{ECOOL2} and accumulated. At the same time
$\simeq 10^{8}$ low energy positrons ($e^{+}$s), emitted by a $^{22}Na$ source,
are slowed down, trapped and accumulated in another trap.\\
One of the major challenges will consist in bringing $\bar{p}$s and $e^{+}$s
in close contact to allow their recombination in a nested Penning trap
\cite{PENN2}. Once $\bar{p}$s and $e^{+}$s are recombined, the confinement by
electric forces ceases and the $\bar{H}$ would escape, hit the nearest wall
and annihilate.\\
To confine the produced $\bar{H}$, the use of magnetic gradients interacting
with the $\bar{H}$'s magnetic moment is investigated for the second phase.

%%%%

\section{The Beam Detector}
The ATHENA beam line is sketched in fig. \ref{fig:beam_line}:
\begin{figure}[h]
 \begin{center}
  \mbox{\psfig{file=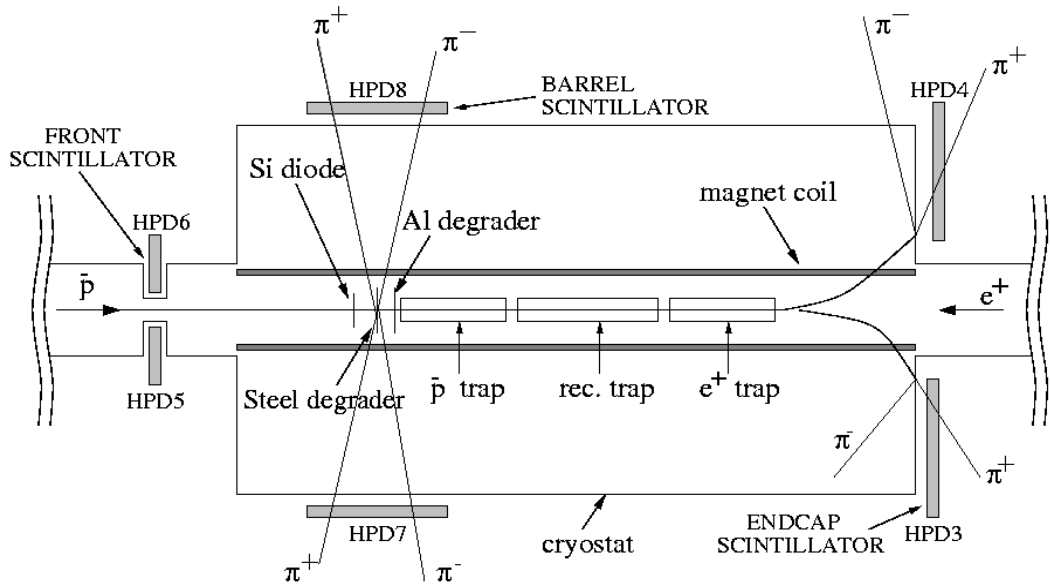,width=15cm}}
  \caption{Schematic upper view of the beam line setup; the scintillators are
           shown together with the name of the HPDs to which they are coupled.
           HPD7 and 8 have a window diameter of 25 mm, while the other ones of
           18 mm (see later in par. 3).}
  \label{fig:beam_line}
 \end{center}
\end{figure}
the injected $\bar{p}$s travel inside a solenoidal magnetic field of 3 T and
are degraded by $25\,\mu m$ of stainless steel put 10 cm in front of the
$\bar{p}$ catching trap and by $130\,\mu m$ of Al put on the first electrode
of the same trap. The degraders further slow down the beam in order to have a
better trapping efficiency; the possibility to use only one gas degrader to
have more flexibility is under study.\\
A $67\,\mu m$ thick Si diode segmented in $5$ pads, located 15.5 cm before
the $\bar{p}$ trap, measures the beam position just before the entrance window
of the trap. The beam transmission of the Si diode is $\simeq 100\,\%$.\\
On the basis of a Monte Carlo simulation, a trapping efficiency of
$\simeq 0.1\%$ is expected with $\simeq 50\%$ of the beam that annihilates on
the degraders and the remaining $\simeq 50\%$ with energy higher than
$(5\div 10)$ keV, corresponding to the highest trapping potential, that
escapes from the trap and travels until the end of the cryostat where it
annihilates.\\
A beam detector is necessary to monitor the beam intensity and the beam
stability, to detect possible losses of $\bar{p}$ along the beam pipe and to
measure the $\bar{p}$ trapping efficiency. Such a device is supposed to detect
the pions coming from the annihilation of the $\bar{p}$s in the degraders, on
the end of the cryostat and, eventually, in the beam pipe. Being the rate of
the pions very high ($\simeq 10^{14}$ Hz), the beam detector can't operate in
pulse mode counting the single particles: hence the current mode operation,
measuring the total charge generated by the pions in the active volume during
every shot, is necessary. To fulfil all these tasks, the beam detector must be
stable and linear to high intensities and over a wide dynamic range;
furthermore a good flexibility is required until the beam features are
completely understood.
\subsection{Scintillator layout and assembling}
The beam detector consists of 3 systems each of 2 modules of plastic
scintillator Bicron BC408 (with peak emission wavelength of 425 nm). The 3
systems, referred to as FRONT, BARREL and ENDCAP, are sketched in
fig. \ref{fig:beam_line} while their dimensions are reported in
table \ref{tab:scint_dim}.

\begin{table}[htb]
 \centering
 \begin{tabular}{|c|c|c|c|}
  \hline
   Scintillator System & Length (cm) &  Width (cm) & Thickness (cm) \\
  \hline
   FRONT               & 20.0        &   10.0      & 1.0            \\
  \hline
   BARREL              & 80.0        &   20.0      & 1.0            \\
  \hline
   ENDCAP              & 39.5        &   20.0      & 1.0            \\
  \hline
 \end{tabular}
 \caption{Dimensions of the 3 scintillator systems.}
 \label{tab:scint_dim}
\end{table}
\par
All the scintillators are glued to lucite light guides from one side, wrapped
with aluminized mylar sheets, sealed with black paper and coupled with optical
grease (BICRON BC-630) to the photo-detector.\\
Every scintillator is equipped with a blue LED (with peak emission wavelength
$423$ nm) glued in a drilled hole in the middle of the scintillator side
facing the light guide. The LED can be driven by a pulser (LeCroy 9212) to
test the photodetector.\\
The ENDCAP scintillators were not used in the first run.
\subsection{Photo-detector choice}
The number of collected photons, produced in a 1 cm thick plastic scintillator
by the charged pions coming from the annihilation of the beam, is given by:
\begin{equation}
 n_{\gamma} = \delta\,n_{\bar{p}}\,\bar{n}_{\pi}\,\Omega\,k\,\epsilon
 \label{eq:int_gamma}
\end{equation}
where $\delta$ is the percentage of dumped beam,
$n_{\bar{p}} \simeq 1.2\times10^{7}$ $\bar{p}$s/shot is the nominal beam
intensity, $\bar{n}_{\pi}=3.5$ is the mean number of charged pions produced in
a $p\bar{p}$ annihilation at rest, $\Omega$ is the solid angle covered by the
scintillator, $k\simeq 1.5\times10^{4}$ $\gamma$/cm$\times$MIP is the mean
number of photons emitted by a $1\,cm$ thick plastic scintillator traversed by
a MIP and $\epsilon$ is the light collection efficiency of the light guide.
Assuming to dump half of the beam in the degraders and assuming
$\Omega \simeq 10\%$ and $\epsilon=50\%$, we obtain
$n_{\gamma}\simeq 1.6\times 10^{10}\,\gamma$/shot: with such an intensity any
photomultiplier tube would saturate for the effect of the high density space
charge. For example, the PMT XP2020 saturates with an intensity of
$\simeq 4\times 10^{5}\,\gamma/200$ ns \cite{XP2020}, around 4 orders of
magnitude below our working intensity. Furthermore the PMTs are very sensitive
to external magnetic fields and their response is a complex function of the
orientation with respect to the field lines.\\
A second choice would be to couple the scintillator directly to a Si photo
diode: assuming a quantum efficiency QE $\simeq 40\%$, the number of
photoelectrons (PEs) would be $n_{PE}\simeq 6.4\times 10^{9}$ PEs and the
corresponding charge $\simeq$ 10 nC (collected in $\simeq$ 250 ns). This
result would be obtained in the best case: if we take into account a realistic
beam variation of a factor 2 in width, of a factor 5 in intensity (both
achievable in the coming years' runs) and a variation in the dumping power
between $10^{-1}$ and 2 (corresponding respectively to a beam transmission of
$5\%$ and to a full dumped beam), we would obtain a variation of a factor $x$,
with $10^{-2} \le x \le 20$. In this situation a preamplifier is required to
detect the low level signals; furthermore a high dynamic range
($\simeq$ 60 dB) is necessary in order to maintain the linearity.\\
The optimal solution would be a Si diode without preamplifier, to not
deteriorate the linearity and the dynamic range of the Si, but with a further
adjustable linear gain of a few hundreds: the proximity focused Hybrid Photo
Diode (HPD).

%%%%

\section{The HPD}
\begin{figure}[h]
 \begin{center}
  \mbox{\psfig{file=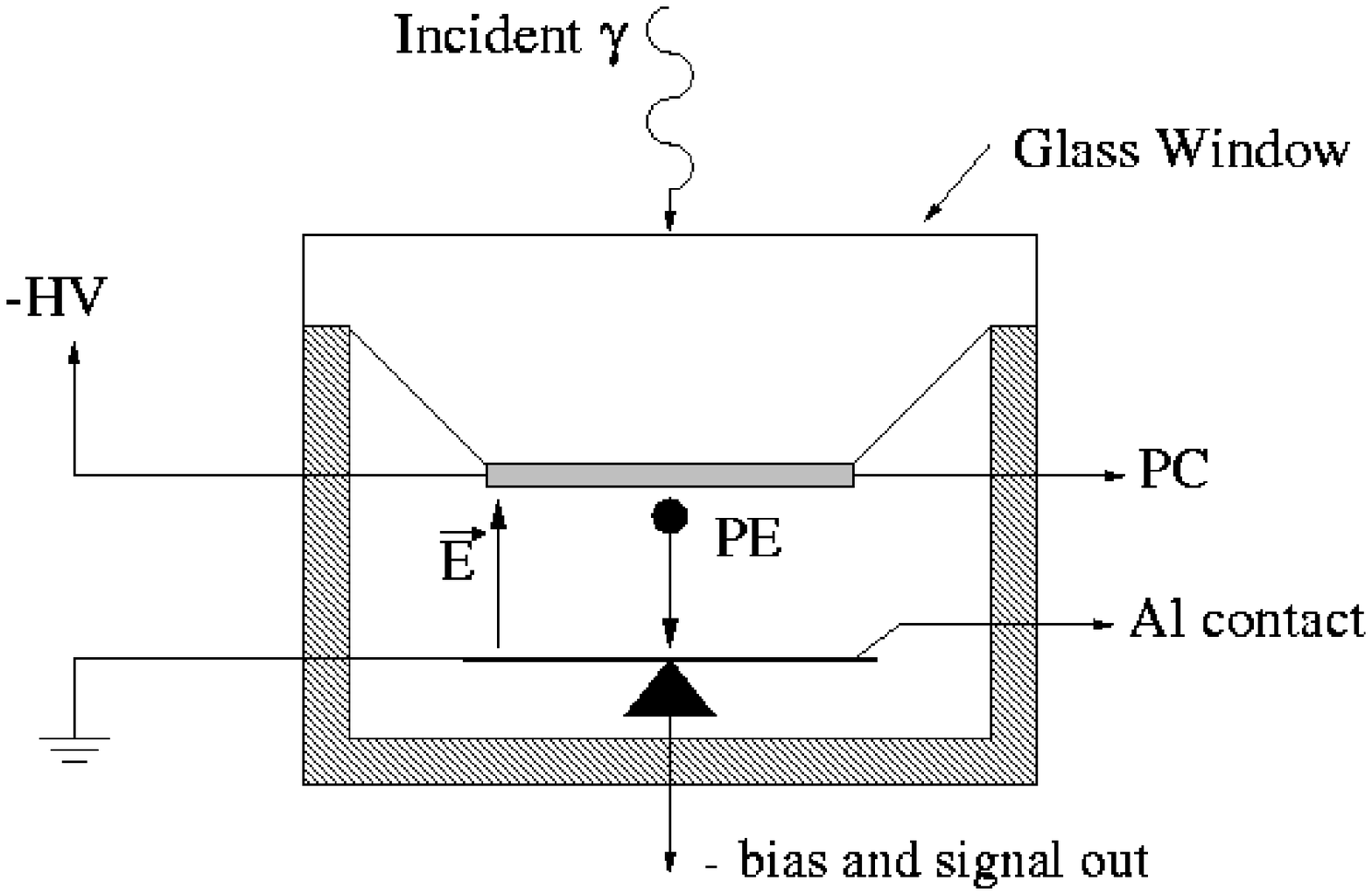,width=8cm}}
  \caption{Schematic working principle of the HPD.}
  \label{fig:HPD_work}
 \end{center}
\end{figure}

The proximity focused HPD \cite{HPD1,HPD2,HPD3,HPD4,HPD5} is a vacuum tube in
which the dynodic ladder is eliminated and the collection electrode is
replaced by a planar Si diode biased in inverse mode. The working principle of
the HPD is schematically shown in fig. \ref{fig:HPD_work}: a PE, extracted
from the photo cathode (PC), is accelerated by a negative high voltage (-HV)
towards a Si chip where it penetrates dissipating its energy creating one
electron-hole (e-h) pair every 3.62 eV of released energy. Since the nature
of the gain is dissipative rather than multiplicative, the HPD turns to be a
linear device with a wide dynamic range and its gain is described by the
following equation:
\begin{equation}
 G = \frac{(HV - V_{th})\,q_{e}}{E_{ion}}
 \label{eq:gain}
\end{equation}
where $V_{th}$ is a voltage threshold, $q_{e}$ is the electron charge and
$E_{ion}=3.62$ eV is the Si ionization energy. The threshold is due to a
passive Al layer deposited on the Si in order to improve the charge collection
and the response uniformity.\\
Applying a HV of $\simeq$ -15 kV a gain of $\simeq$ 3500 can be achieved: this
is low if compared to that of a PMT ($\simeq 10^{7}$) and a further
amplification is necessary in common applications where low light yields are
available \cite{HPD6,LHCb,CMS}. In this application, for which a large amount
of light is available, the idea is to use a HPD, not preamplified, to have a
wide dynamic range.\\
We have used two different models of proximity focused HPDs built by
DEP \cite{DEP}: one for the FRONT and the ENDCAP scintillators (model PP0350F)
and the other for the BARREL scintillators (model PP0350D).
In table \ref{tab:HPD} are reported the main characteristics of the two models.
\begin{table}[htb]
 \centering
 \begin{tabular}{|c|c|c|c|c|}
  \hline
   Model   &  Window Diameter &  Max HV & Capacitance & Max Bias \\
  \hline
   PP0350F &     18. mm       &  -15 kV &   120  pF   &  -130 V  \\
  \hline
   PP0350D &     25. mm       &   -8 kV &   200  pF   &  -90 V   \\
  \hline
 \end{tabular}
 \caption{Main characteristics of the used HPDs.}
 \label{tab:HPD}
\end{table}
\par
In both the models an S20-UV PC with a QE of $20\%$ at 423 nm is deposited on
the window in front of a $300\,\mu m$ thick Si PiN diode in the E-type
configuration.\\
The 18 mm diameter HPDs are the same used in the FINUDA experiment: a full
description of their characterisation can be found in ref. \cite{HPD6}; the
$25$ mm diameter HPDs have been chosen for their bigger surface, in order to
improve the light collection efficiency.
\subsection{Static characterisation}
\begin{figure}[h]
 \begin{center}
  \mbox{\epsfig{file=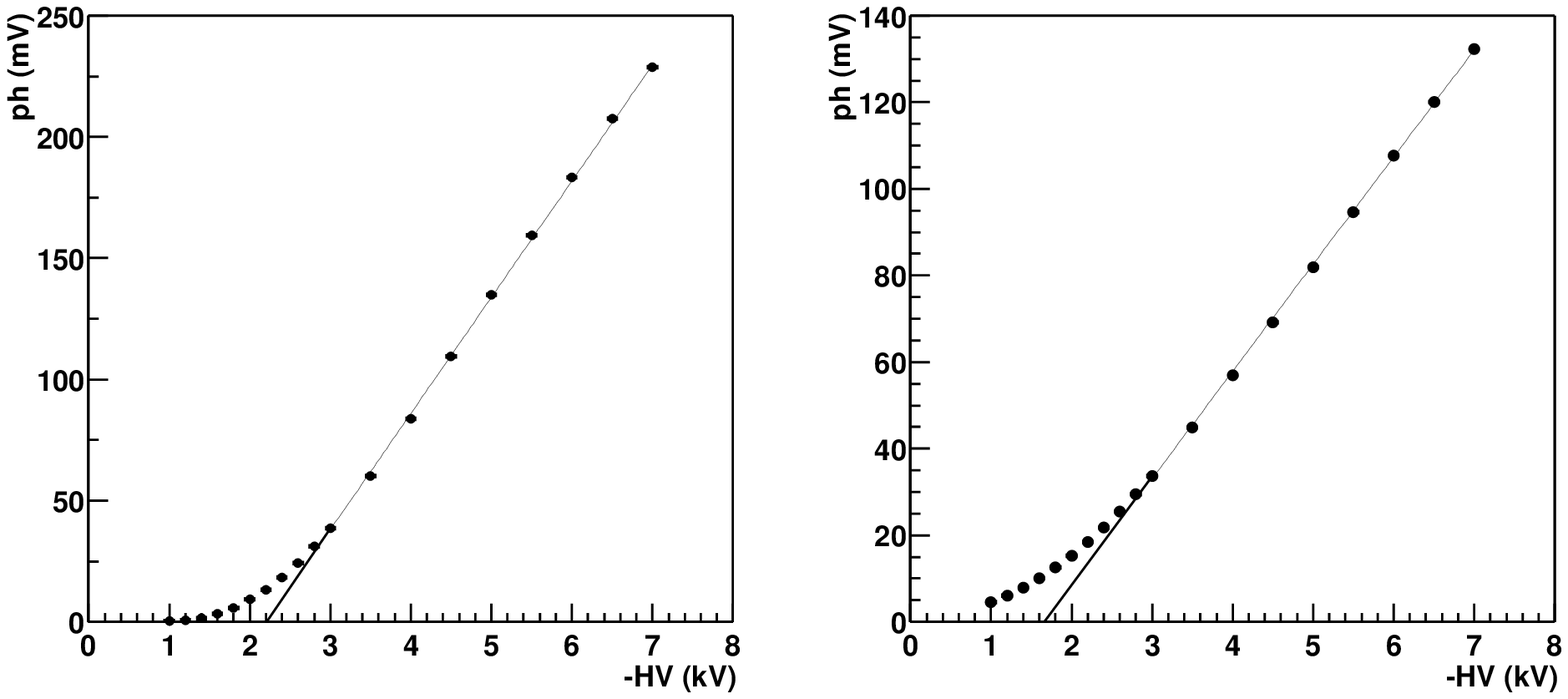,width=14cm}}
  \caption{Signal pulse height versus high voltage for HPD5 (left) and HPD7
           (right). The intercepts with the x axis gives the threshold
           $V_{th}$.}
  \label{fig:thres_13}
 \end{center}
\end{figure}

For each HPD the threshold $V_{th}$ has been measured pulsing the LED with a
fixed current and recording the signal pulse height corresponding to different
HV values. In fig. \ref{fig:thres_13} are reported the results for HPD5 (left)
and HPD7 (right): the experimental points have been fitted with a straight
line from -3.0 kV to -7.0 kV whose crossing point with the x axis gives the
threshold ($V_{th}$).
From the fit we found $V_{th}=(-2.21\pm 0.05)$ kV for HPD5 and
$V_{th}=(-1.66\pm 0.05)$ kV for HPD7. Once the thresholds are known, the gain
can be rescaled for different HV operating values.\\
The noise of the HPDs coupled to the scintillators is 0.4 mV (rms),
corresponding to 2 pC. Since signals ranging from a few hundreds of mV to a
few tens of V are expected, all the measurements can be considered with
virtually no noise.
\subsection{Behaviour in an external magnetic field}
No response variation is expected if the magnetic field and the electric field
generated by the HV are parallel, otherwise the PEs are accelerated along a
spiral which loops around the magnetic field lines. The PEs still reach the
Si surface with full energy but in a displaced position and with variable
impinging angle: no signal cutoff is expected unless the PEs are dragged
outside the Si chip. As the impinging angle increases the dead contact layer
on the Si chip becomes difficult to penetrate and the signal is expected to
decrease with increasing magnetic field.\\
Intensive studies of behaviour of a proximity focused HPD in an external
magnetic fields have been performed \cite{HPD4}.\\
We measured the response of the HPDs in the ATHENA magnetic field using a
pulsed LED, hold in front of the PC's centre by a plastic support fixed to
the HPD. We also measured the field intensity with a portable gauss-meter.\\
A great care was taken in choosing a position of the HPDs so that the field
intensity was less than 1 kG and the angle with respect to the HPD electric
field was less than 30$^{o}$ (though it was not easy to measure with good
precision the intensity and the direction of the field) where no losses are
foreseen \cite{HPD4}. The pulse height of the signal extracted from the HPDs
was measured with a digital scope (LeCroy 9354A). The measurement was
repeated with the magnetic field switched off: the difference between the two
measurements was compatible with zero within $3\%$. The difference are mainly
due to the LED movements during the measurements, due to the home made
setup.\\
Also the threshold was measured with and without magnetic field and no
variation was found within the errors ($3\%$).
\subsection{Power supply}
\begin{figure}[h]
 \begin{center}
  \mbox{\psfig{file=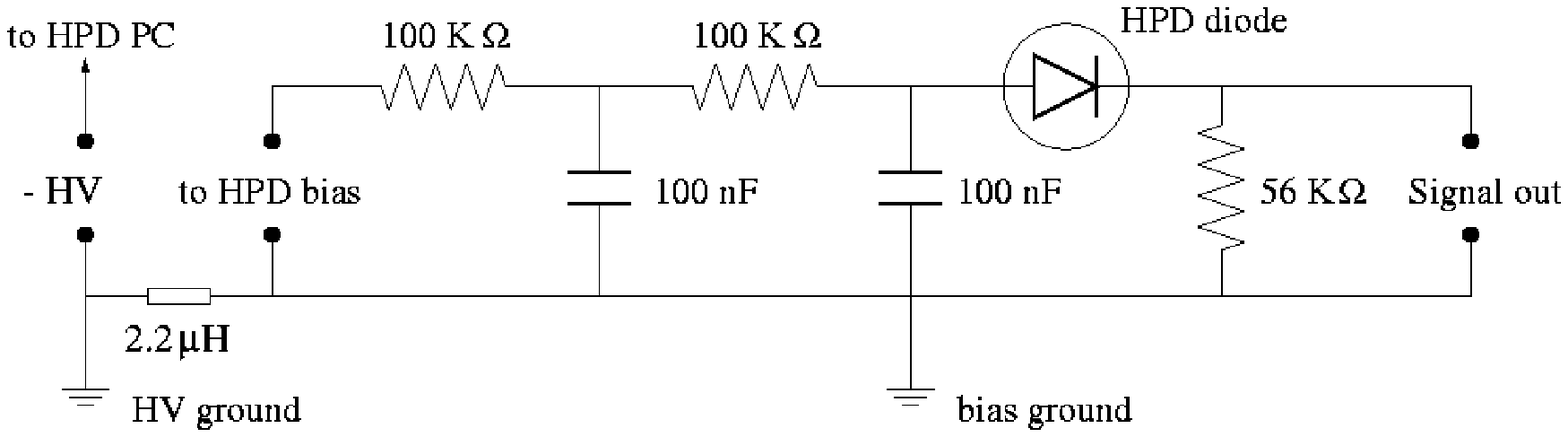,width=15cm}}
  \caption{Wire diagram of the bias supply circuit.}
  \label{fig:bias_schema}
 \end{center}
\end{figure}

The HV from the power supply (Heinzinger HNC 30 kV-20 mA) has been filtered by
a $\pi$-filter with two capacitors of 100 nF and a resistor of 100 k$\Omega$
and then sent to the PC through a $4\,m$ long coaxial cable (Nokia
Kabel-HTC-50-2-1 FRNC 95/30). A big care was taken in the insulations of the
high voltage contacts, mostly performed with teflon shields.\\
The bias voltage of each HPD diode has been filtered and distributed by a
small bias circuit, sketched in fig. \ref{fig:bias_schema}, mounted on the
back side of each HPD.

%%%%

\section{First results}
\begin{figure}[h]
 \begin{center}
  \mbox{\psfig{file=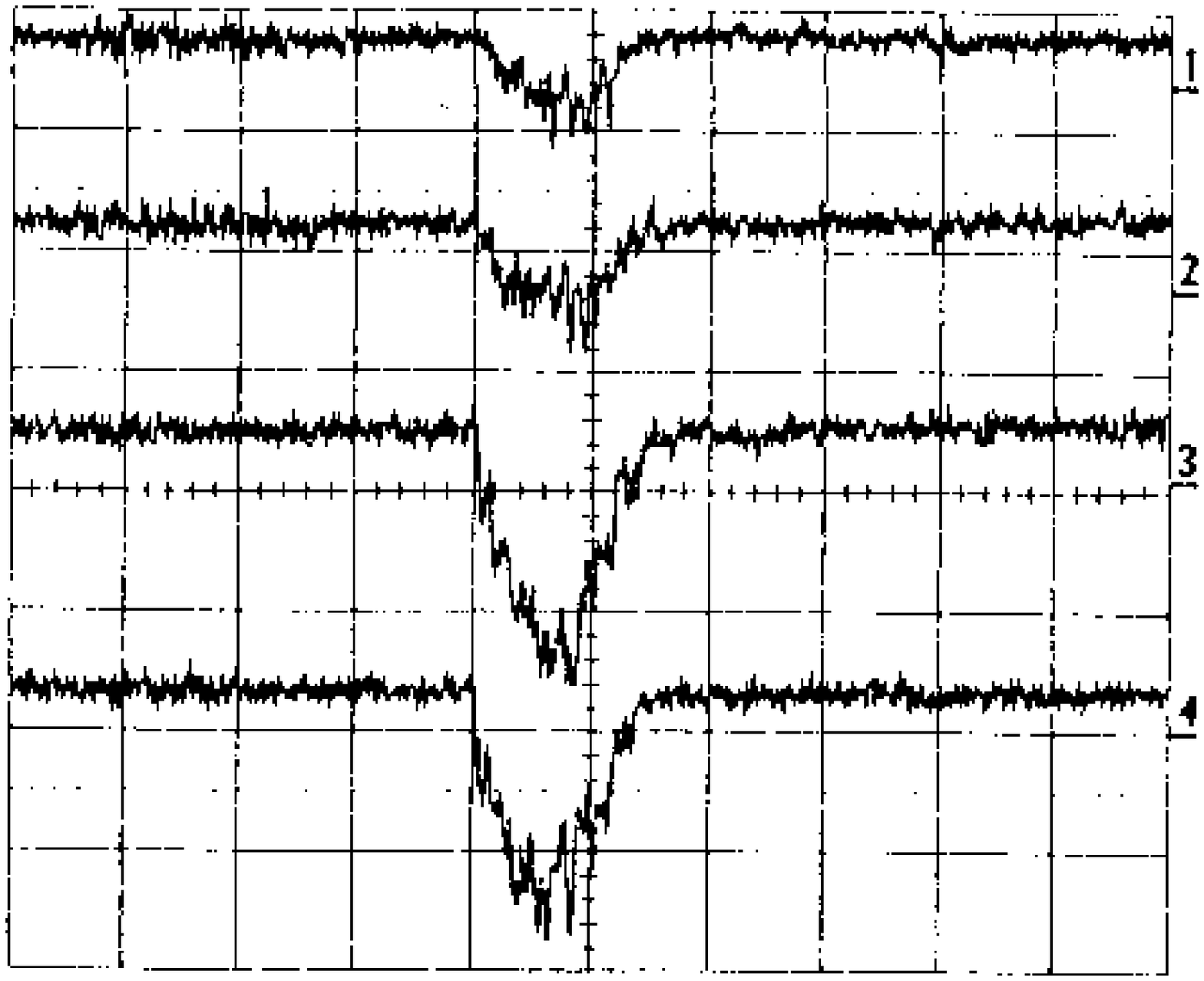,width=7cm}}
  \caption{First signals detected by the beam counter prototype. The
           horizontal pitch is 500 ns/division and the vertical one is
           2 mV/division.}
  \label{fig:first_burst}
 \end{center}
\end{figure}

AD delivered the first $\bar{p}$s from November 30-th, 1999 for 3 days: during
this run the electron cooling was not optimised and the beam intensity was
$\simeq 10^{3}\,\bar{p}$s/shot, 3 to 4 orders of magnitude reduced compared to
the nominal one.\\
A prototype of the beam detector was tested to check its feasibility and its
reliability: it consisted of 2 BARREL scintillators, each coupled to both
sides to two $18$ mm diameter HPDs, one (BARREL1) aligned with the Al
degrader, the other (BARREL2) positioned $\simeq 2$ m downstream far from the
$\bar{p}$ injection region. The signals extracted from the HPDs were directly
fed into a digital scope (LeCroy model 9354A) triggered by the AD extraction
signal. The HV was set to -15 kV and the bias to -120 V.\\
In fig. \ref{fig:first_burst} are shown some typical signals from
the HPDs: the waveforms 3 and 4 come from the BARREL1 scintillator, the
waveforms 1 and 2 from the BARREL2.
The attenuation of the signals 1 and 2 with respect to 3
and 4, is due to the low $\bar{p}$ transmission of the degrader. All the
signals are well correlated in time: the bunch length is $\simeq 500$ ns
(FWHM). The charge collected by the HPDs 3 and 4 is $\simeq 40$ pC,
corresponding to $3.6\times 10^{2}\,\bar{p}$s, in accordance with the beam
intensity if we take into account that part of the beam was lost in the beam
pipe because of the beam blow up for the not optimised electron cooling.\\
From these measurements we deduced that the HPD in this configuration without
preamplifier could be used as photo detector for the beam counter and that it
was a flexible device: in fact we were able to detect a signal a factor
$\simeq 10^{3}$ lower than the nominal one, increasing the HPD gain by the
same amount.

%%%%

\section{Relative calibration and trapping efficiency}
From a Monte Carlo simulation the best trapping efficiency ($\simeq 0.1\%$) is
obtained when half of the beam annihilates on the degraders and the remaining
escapes the trap and annihilates at the end of the cryostat. The trapping
efficiency can be optimised and monitored very precisely in future with a
variable degrader. The beam will be fully stopped: in this case the ratio
between the charge measured by the BARREL system over the ENDCAP will show a
maximum; the opposite situation will be measured with a fully transmitted
beam. These measurements will give the two extreme points of a full dumped
beam and of a full transmitted beam: the working point, corresponding to the
best trapping efficiency, will be most likely in the middle. It must be
stressed that no absolute calibration is necessary for this measurement,
because only charge ratios are involved. A fine tuning can be done by dumping
the trapped $\bar{p}$s on the Al degrader by inverting the trap potential and
then counting the charged pions from their annihilation with a scintillator
coupled to fast PMTs. Once the best settings are found, the intensity can be
monitored in parasitic way shot by shot as described before.

%%%%

\section{Absolute calibration}
\begin{figure}[h]
 \begin{center}
  \mbox{\psfig{file=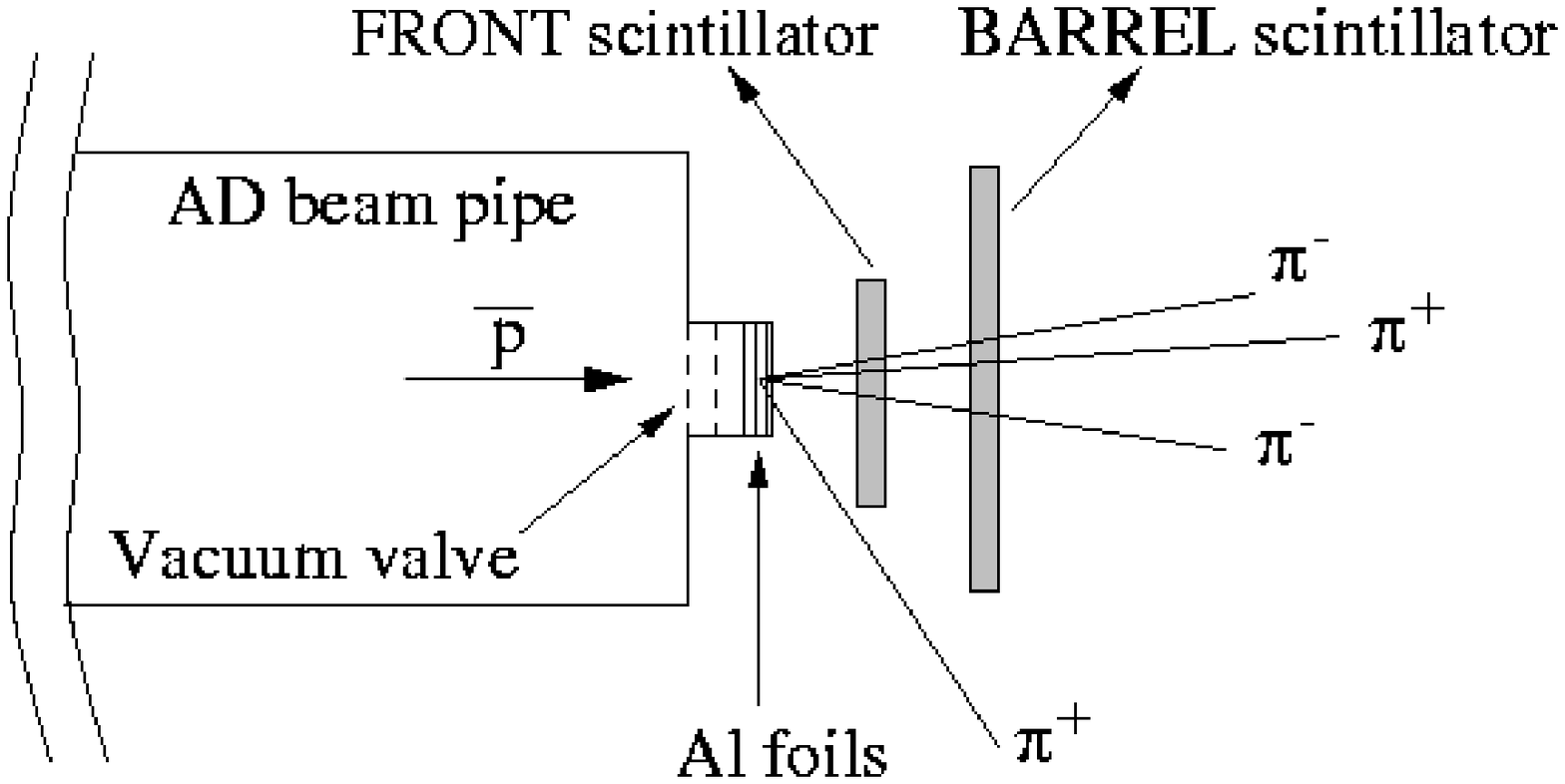,width=11cm}}
  \caption{Schematic upper view of the setup used for the absolute
           calibration.}
  \label{fig:beam_stopped_Al}
 \end{center}
\end{figure}

The beam detector cannot be calibrated with a radiation source, because the
HPD, in this configuration, is not sensitive to single particle counting.\\
An absolute calibration has been performed with an Activation Method (AM)
and a Schottky Method (SM), described in the following, using the experimental
setup sketched in fig. \ref{fig:beam_stopped_Al}.
\par
\begin{figure}[h]
 \begin{center}
  \mbox{\psfig{file=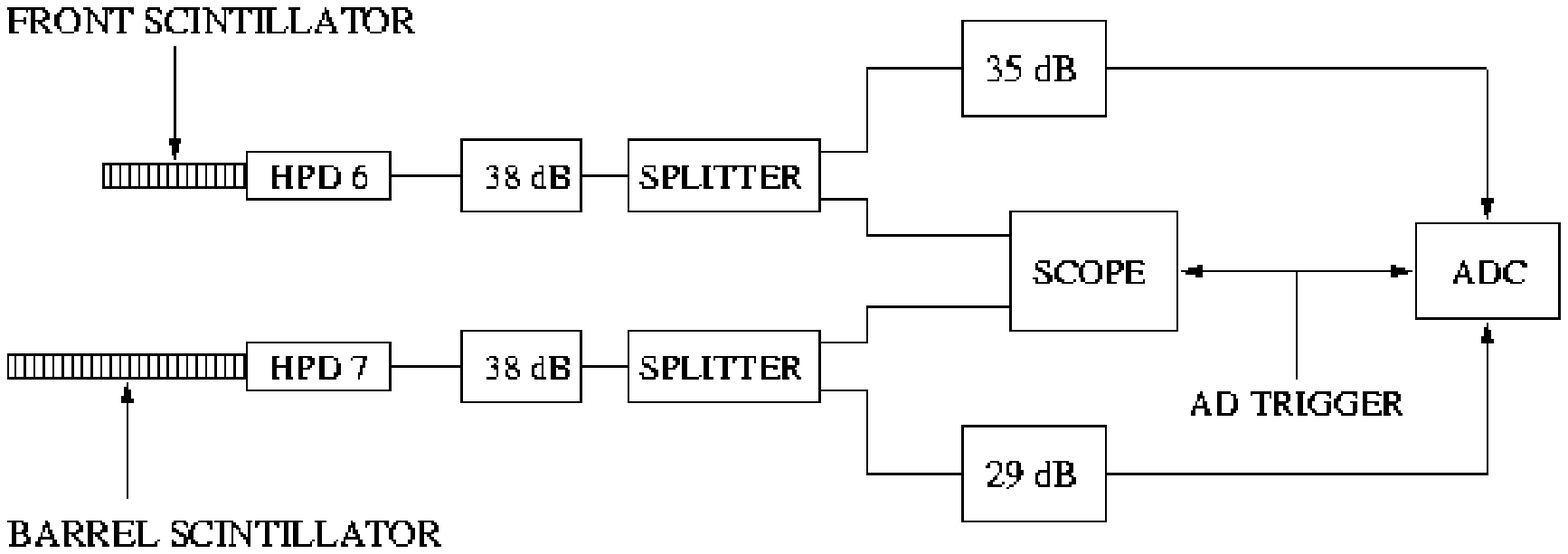,width=11cm}}
  \caption{Layout of the DAQ chain used for the absolute calibration.}
  \label{fig:DAQ_Al}
 \end{center}
\end{figure}

Two scintillators were calibrated: a FRONT one (coupled to HPD6), and a
BARREL one (coupled to HPD7) positioned with their centres aligned with the
beam line; the solid angles covered by the FRONT and the BARREL scintillators
are respectively $\Omega_{FRONT}=23.7\%$ and $\Omega_{BARREL}=24.1\%$.\\
The HV of the HPDs was set to -2.5 kV and the bias to -70 V.\\
In fig. \ref{fig:DAQ_Al} is reported the layout of the DAQ chain used for the
test.
\par
The signal from the HPDs was attenuated by 38 dB and then split with a linear
fan-in-fan-out: one end was sent to the scope, the other one was further
attenuated by 35 dB for HPD6 and by 29 dB for HPD7 and finally fed into an ADC
(LeCroy model 1182) with a charge resolution of 50 fC. Both the scope and the
ADC were triggered by the AD warning signal.\\
The ADC data were transfered to a personal computer via a VME/MXZ bus, while
the scope data were readout via GPIB and both recorded on a disk. The number
of $\bar{p}$s was measured for each shot before the extraction by a low noise
Schottky pickup (SP) \cite{SP1,SP2}, with a sensitivity of $\simeq 1\%$.\\
The beam was fully dumped by an Al foil, positioned at the entrance of the
ATHENA beam pipe, as shown in fig. \ref{fig:beam_stopped_Al}. A typical signal
from HPD7 acquired with the scope is reported in fig. \ref{fig:sigHPD7_Al}:
\begin{figure}[h]
 \begin{center}
  \mbox{\epsfig{file=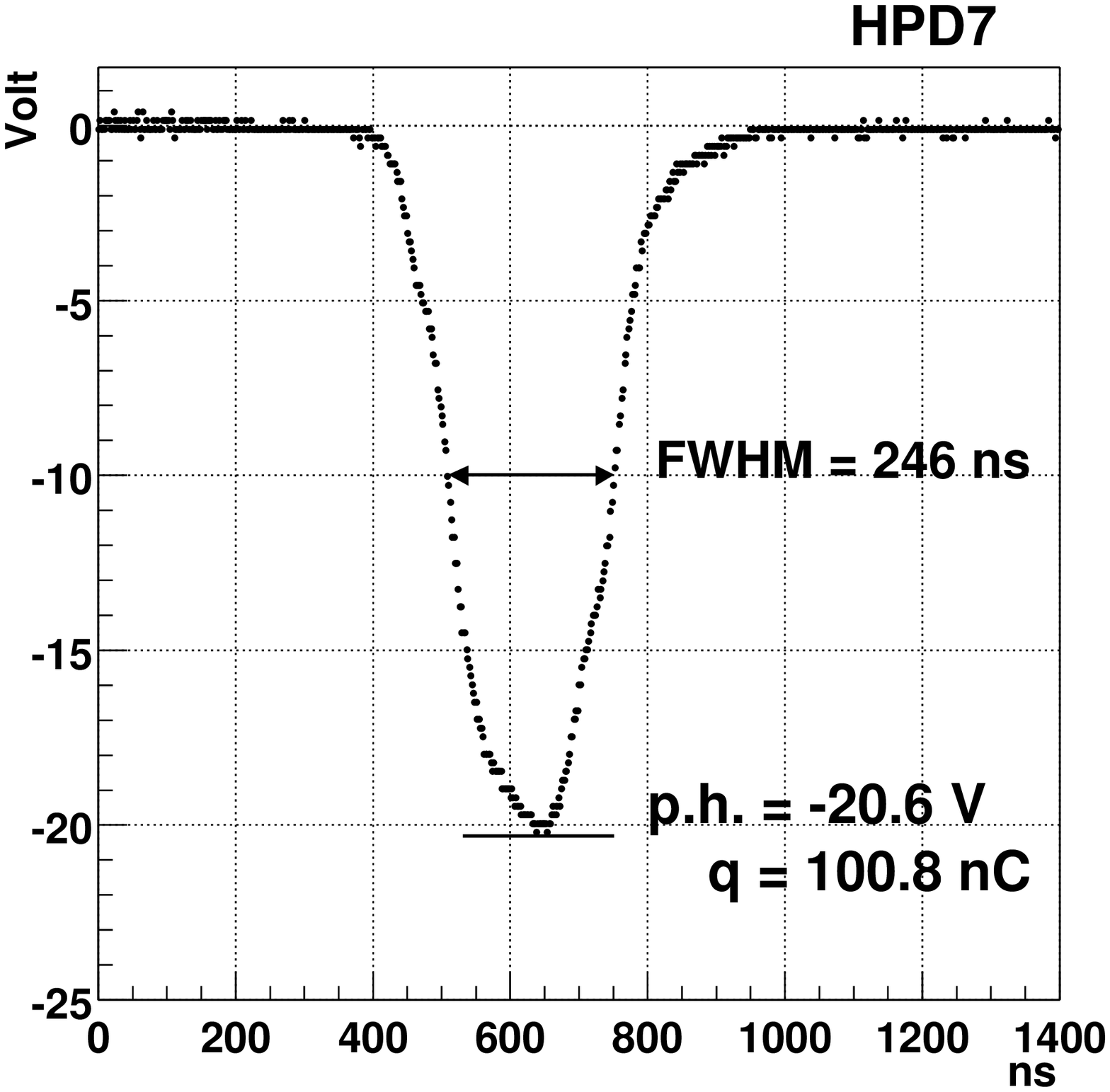,width=7cm}}
  \caption{Typical waveform from HPD7.}
  \label{fig:sigHPD7_Al}
 \end{center}
\end{figure}
the pulse height is -20.60 V (over 50 $\Omega$), the collected charge is
100.84 nC and the width is 246.1 ns (FWHM). The waveform follows accurately
the longitudinal bunch shape generated by the RF system.\\
A total of 32 shots was acquired in 1h 46s: for 3 of these, no beam was
ejected from AD, for other 9 the number of ejected $\bar{p}$s was not recorded
due to a problem in the Schottky DAQ; for these events the number of
$\bar{p}$s was extrapolated from the charge by rescaling it with respect to
the other ones.

\begin{figure}[h]
 \begin{center}
  \mbox{\epsfig{file=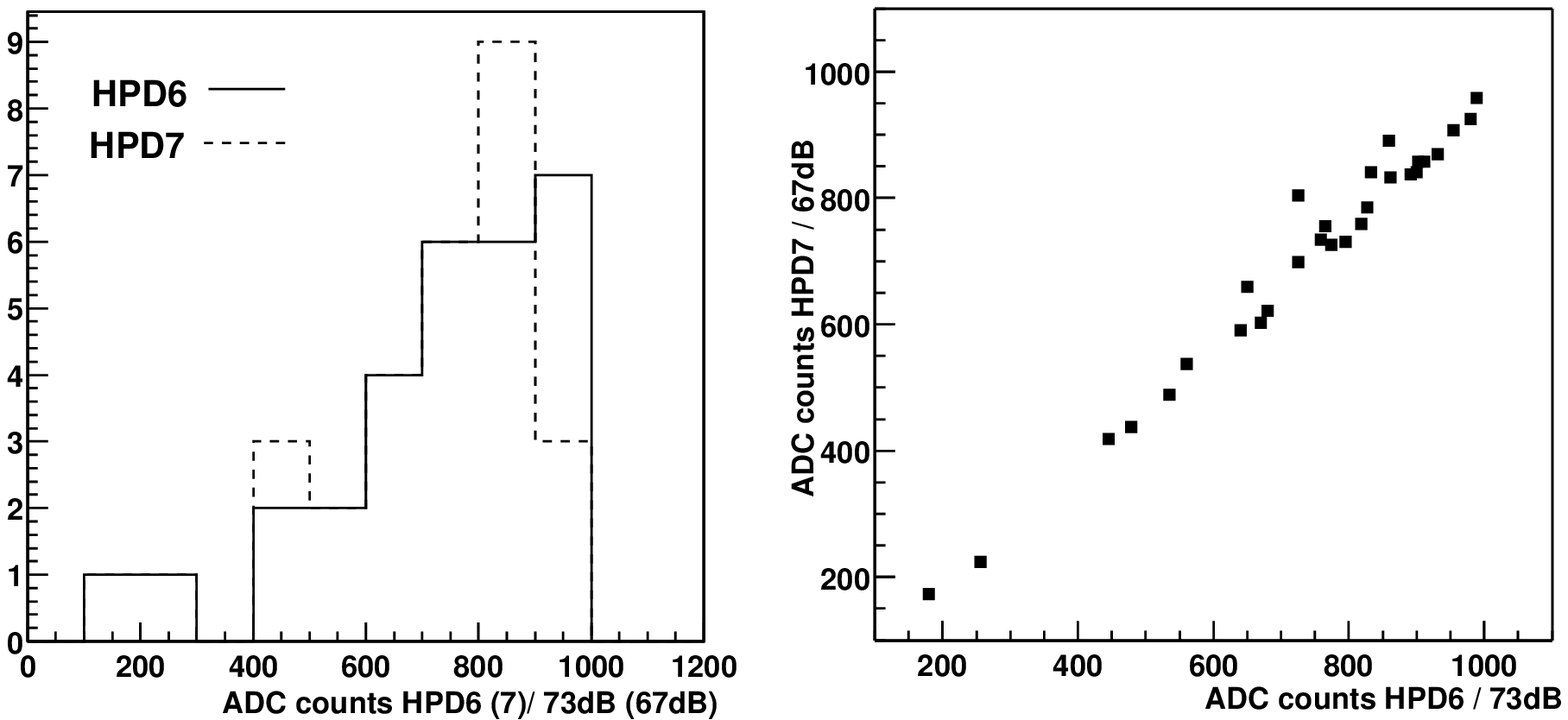,width=14cm}}
  \caption{Histograms of the ADC counts (divided by the attenuation factors)
           of HPD6 (left, solid line), HPD7 (left, dashed line) and their
           correlation (right).}
  \label{fig:nadc_Al}
 \end{center}
\end{figure}

In fig. \ref{fig:nadc_Al} the histograms of the ADC counts (divided by the
attenuation factors) of HPD6 (left, solid line), of HPD7 (left, dashed line)
and of their correlation (right) are reported.
\par
The mean number of ADC counts (corrected for the attenuation factors) is
$(3.28 \pm 0.07)\times 10^{6}$ ADC counts for HPD6 and
$(1.57 \pm 0.03)\times 10^{6}$ ADC counts for HPD7.
\subsection{Activation Method}
This method relies on the following activation process, studied in LEAR
\cite{Activ}:
\begin{eqnarray}
 \bar{p} + Al \rightarrow ^{24}Na^{*}
 \label{eq:Al_activ1}
 \\
 ^{24}Na^{*} \rightarrow ^{24}Na^{*} + \gamma
 \label{eq:Al_activ2}
\end{eqnarray}
The branching ratio of reaction (\ref{eq:Al_activ1}) is $BR=(2.1\pm 0.3)\%$,
while that of reaction (\ref{eq:Al_activ2}) is $BR=100\%$. The half life of
$^{24}Na^{*}$ is $T_{\frac{1}{2}} = 15$ h and the $\gamma$ ray can be emitted
with two different energies: $E_{1} = 1369$ keV and $E_{2} = 2754$ keV.\\
The basic idea is to dump all the beam extracted from the AD on an Al foil and,
after the irradiation, to measure the induced activity counting the $\gamma$s
from reaction (\ref{eq:Al_activ2}) from which the beam intensity can be
derived.\\
Four Al sheets, each with a thickness of $110\,\mu m$, were sticked together
and placed in vacuum; other sheets were placed around the valve to measure the
background.\\
The activity induced by the beam in the Al sheets, was measured with a Ge
detector at a distance of 4.0 cm and at 24.0 cm from the Al foil and then
extrapolated to the end of the irradiation time: the results are reported in
table \ref{tab:Al_meas_activ}.

\begin{table}[h]
 \centering
 \begin{tabular}{|c|c|c|}
  \hline
                     &  Distance (cm) &   Activity (Bq)     \\
  \hline
    Measurement 1    &   $4.0$        &  $84.0 \pm 4.7$   \\
  \hline
    Measurement 2    &  $24.0$        &  $92.7 \pm 7.4$   \\
  \hline
    Background       &   $4.0$        &  $0.0$              \\
  \hline
 \end{tabular}
 \caption{Measured activities of the Al foil extrapolated to the end of
          the irradiation.}
 \label{tab:Al_meas_activ}
\end{table}
\par
The beam intensity can be derived from the measured activity making use of the
following equation \cite{Activ}:
\begin{equation}
\sum_{i=0}^{N} A_{i} =
\sum_{i=0}^{N} \frac{n_{i}^{\bar{p}} P}{\tau}\,e^{-\frac{t_{M}-i\bar{t}}{\tau}}
 \label{eq:activ_2_npbar}
\end{equation}
where $N=32$ is the number of shots, $A_{i}$ is the activity induced in the Al
foil by the $i$-th shot, $n^{\bar{p}}_{i}$ is the number of $\bar{p}$s of the
$i$-th bunch, $t_{M}=$1h 46s is the duration of the irradiation,
$\bar{t}=$118 s is the inter-bunch time, $P=(2.1 \pm 0.3)\%$ is the activation
probability of reaction (\ref{eq:Al_activ1}), $\tau$ is the mean lifetime of
$^{22}Na^{*}$.\\
Assuming a uniform activation induced by every shot, approximating
$e^{\frac{i\bar{t}}{\tau}} \simeq 1+\frac{i\bar{t}}{\tau}$ and taking into
account that the shots number 18, 19 and 30 came without beam, the integrated
number of $\bar{p}$s, $n_{\bar{p}}^{tot}=\sum_{i} n_{i}^{\bar{p}}$, and the
mean number of $\bar{p}$s per shot have been calculated and the results are
reported in table \ref{tab:Al_res_activ}.
\begin{table}[htb]
 \centering
 \begin{tabular}{|c|c|c|}
  \hline
      Activity        & Integrated number of $\bar{p}$s & Mean number of $\bar{p}$s/shot \\
  \hline
   $(84.0\pm 4.7)$ Bq &     $(3.416\pm 0.180)\,10^{8}$    & $(1.101\pm 0.062)\,10^{7}$  \\
  \hline
   $(92.7\pm 7.4)$ Bq &     $(3.523\pm 0.281)\,10^{8}$    & $(1.215\pm 0.097)\,10^{7}$  \\
  \hline
   Combined           &     $(3.477\pm 0.244)\,10^{8}$    & $(1.199\pm 0.084)\,10^{7}$  \\
  \hline
 \end{tabular}
 \caption{Integrated number of $\bar{p}$s and mean number of $\bar{p}$s/shot
          calculated from the corresponding measured activity.}
 \label{tab:Al_res_activ}
\end{table}
\par
The combined result is the weighted mean of the measurements taken at
the two different distances  weighted with the distance.
\subsection{Schottky Method}
This method relies on the direct measurement of the number of $\bar{p}$s
inside the AD ring before the electron cooling at 100 MeV/c and before the
ejection by means of the SP \cite{SP1}. The electron cooling efficiency has
been measured by rebunching the beam in harmonic 3 (where the SP is more
sensitive at 100 MeV/c) before the ejection and it was $\simeq 100\%$.
The total number of $\bar{p}$s is $(3.618 \pm 0.114)\,10^8$, while the mean
number of $\bar{p}$s per shot is $(1.248 \pm 0.039)\,10^7$.
\subsection{Comparison between the AM and the SM}
The calibration factors obtained with the AM and with the SM (with the ADC
counts normalised to the solid angle) are reported in
table \ref{tab:Al_compar}.
\begin{table}[htb]
 \centering
 \begin{tabular}{|c|c|c|}
  \hline
   Calibration Method    & HPD6 (ADC counts/$\bar{p}$) & HPD7 (ADC counts/$\bar{p}$)  \\
  \hline
   Activation (84.0 Bq)  &       $1.257 \pm 0.076$       & $0.592 \pm 0.035$   \\
  \hline
   Activation (92.7 Bq)  &       $1.139 \pm 0.094$       & $0.536 \pm 0.044$   \\
  \hline
   Activation (combined) &       $1.154 \pm 0.084$       & $0.543 \pm 0.039$   \\
  \hline
   Schottky              &       $1.109 \pm 0.042$       & $0.522 \pm 0.019$   \\
  \hline
 \end{tabular}
 \caption{Calibration factors obtained with the AM and SM.}
 \label{tab:Al_compar}
\end{table}
\par
The results are compatible within the errors and show that an absolute
calibration is possible at the level of 7.3$\%$ for the AM and of 3.8$\%$ for
the SM. The calibration performed with the SM is twice more accurate, being
the sensitivity of the SP very high ($\simeq 1\%$) at these intensity. On the
other hand the number of $\bar{p}$s is measured inside the ring before the
electron cooling process at 100 MeV/c and before the extraction and so
possible cooling inefficiencies and losses in the ejection line are not taken
into account. For these reasons the AM is more reliable than the SM, at least
in this experimental situation, because the number of $\bar{p}$s is measured
at the end of the ejection line, even if with a worse error. A better
calibration with the SM will be possible next year when the ejection line
will be equipped of a low noise SP to measure directly the beam intensity
after the extraction.

%%%%

\section{Full dumped beam test}
\begin{figure}[h]
 \begin{center}
  \mbox{\psfig{file=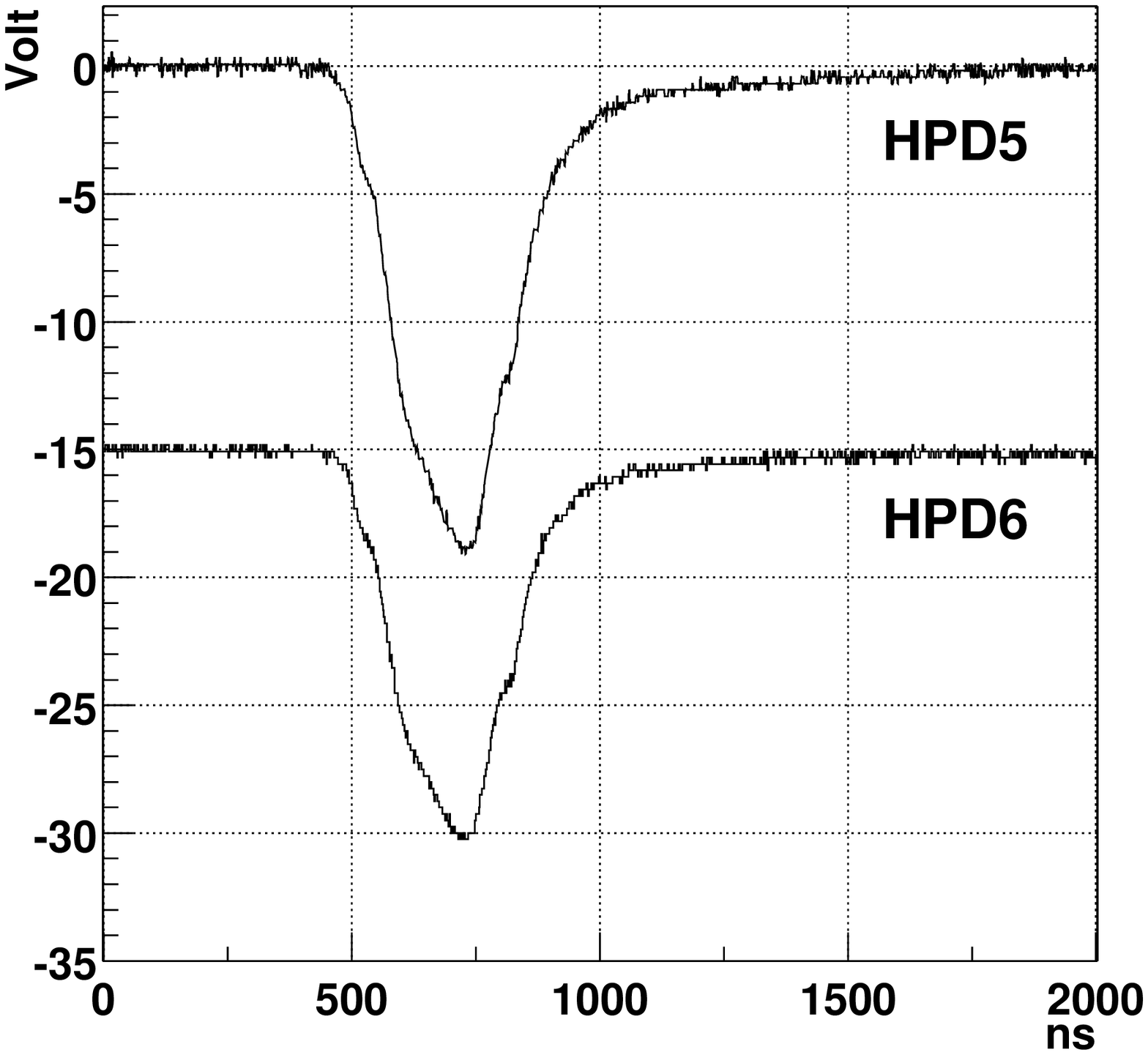,width=6.8cm}}
  \mbox{\psfig{file=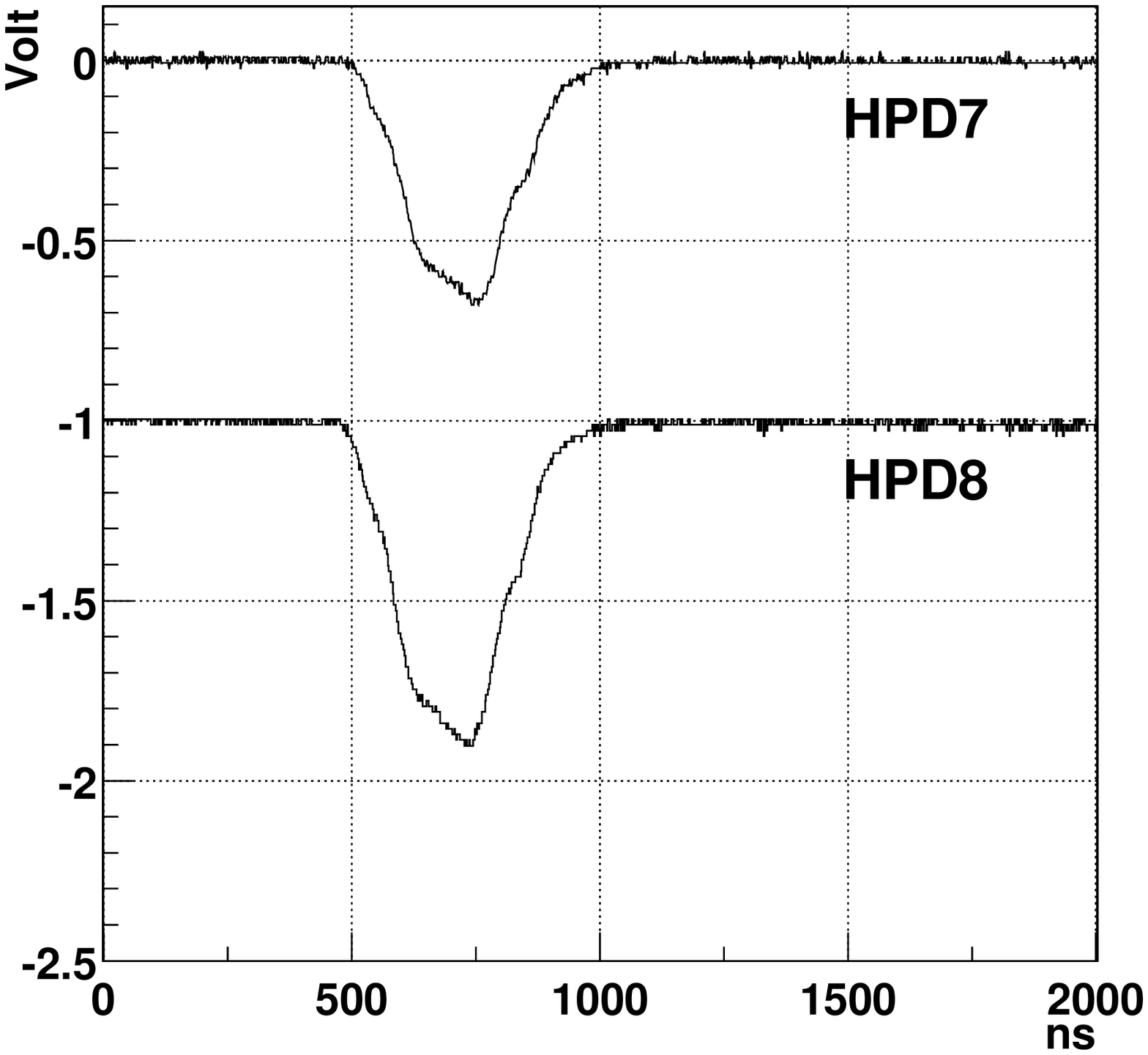,width=6.8cm}}
  \caption{Signals extracted from the HPDs with the beam completely dumped.}
  \label{fig:fullstop_sigls}
 \end{center}
\end{figure}

\begin{figure}[h]
 \begin{center}
  \mbox{\psfig{file=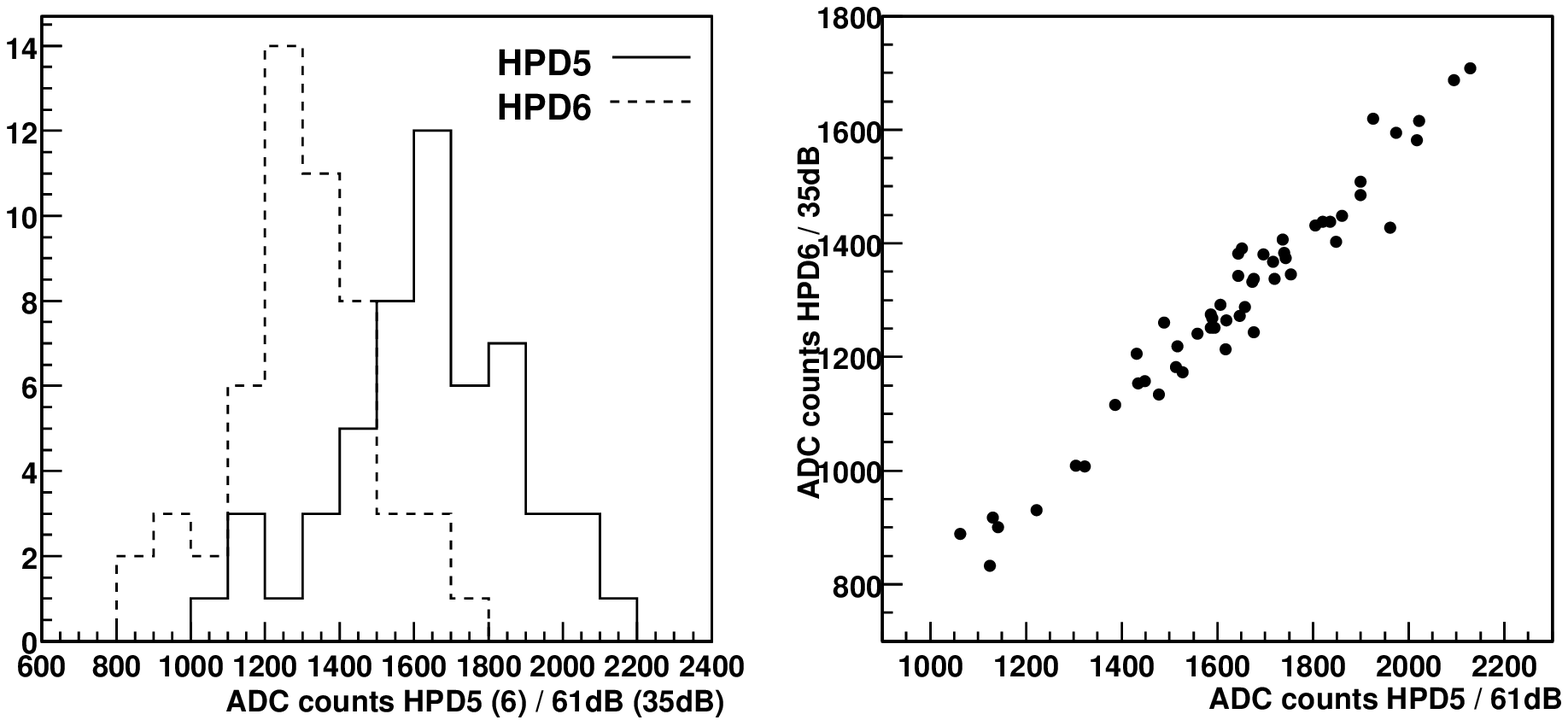,width=14cm}}
  \caption{Histograms of the ADC counts (divided by the attenuation factors)
           from HPD5 (left, solid line), HPD6 (left, dashed line) and their
           correlation (right).}
  \label{fig:fullstop_adc56}
 \end{center}
\end{figure}

A second test with the beam completely dumped was performed to measure the HPD
response to a large amount of light and to perform an absolute calibration
with the SM. The used experimental set up is shown in fig.
\ref{fig:fullstop_setup}:
\begin{figure}[h]
 \begin{center}
  \mbox{\psfig{
        file=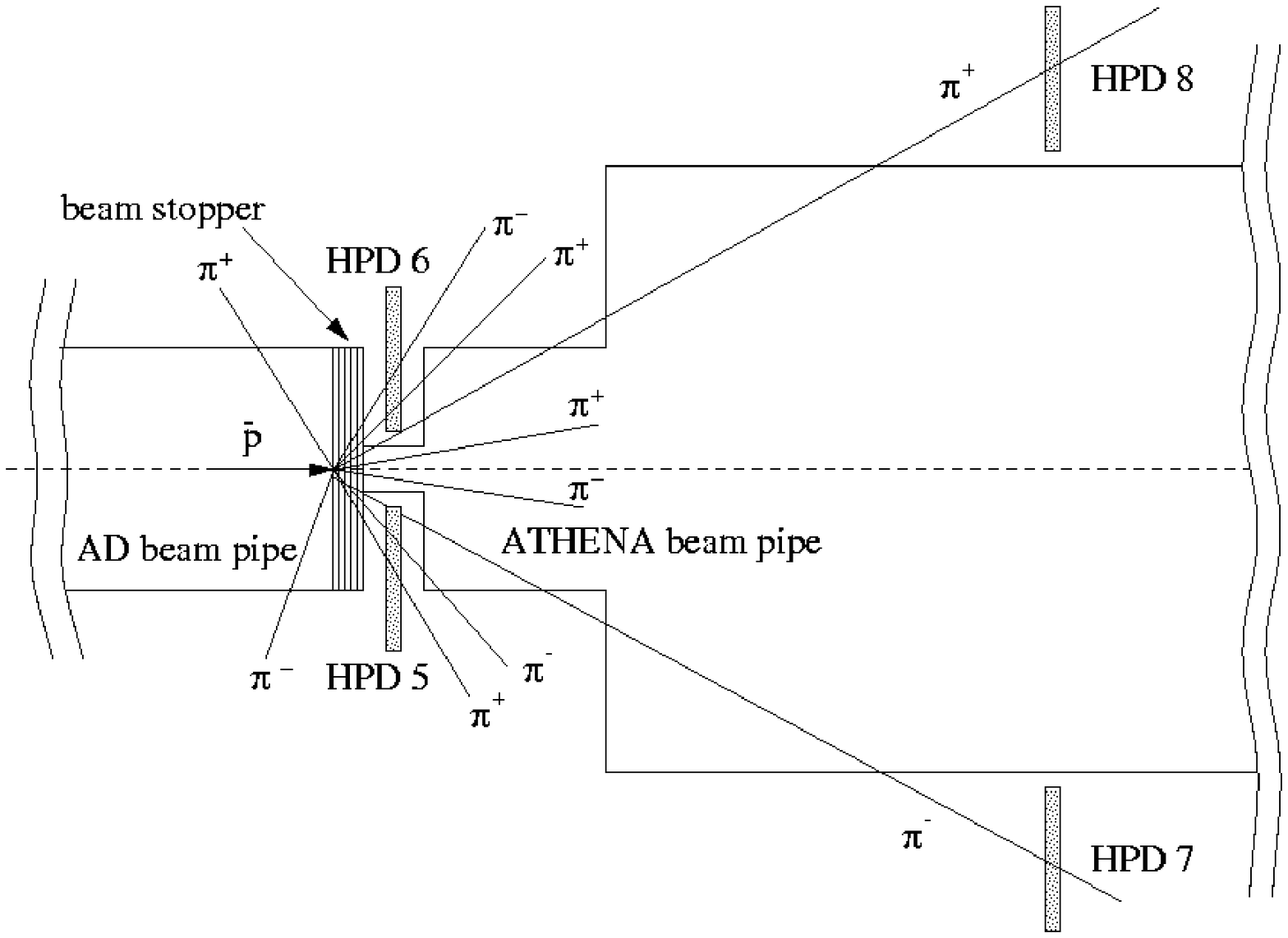,width=11cm}}
  \caption{Schematic upper view of the setup used for the full stopped beam
           tests.}
  \label{fig:fullstop_setup}
 \end{center}
\end{figure}
the FRONT scintillators, coupled to HPD5 and HPD6, and the BARREL ones,
coupled to HPD7 and HPD8 were used. The solid angles covered by a FRONT
scintillator is $6.2\%$, that by a BARREL one is $0.49\%$. The PC high voltage
was set at -4 kV and the bias at -80 V; the signals from the HPDs were
attenuated by a total of 61 dB for HPD5 and HPD6, and by 35 dB for HPD7 and
HPD8.\\
In fig. \ref{fig:fullstop_sigls} are reported the signals directly extracted
from the diode's output and sent to the scope: they are correlated in time and
follow the longitudinal beam structure. The long tail on the trailing
edge of the signals from HPD5 and HPD6 maybe due to the screening of the bias
field by the large charge generated in the Si chip (plasma effect). The tail
effect is of $\simeq 10\%$ and all the charge is completely collected in
$\simeq 1.5\,\mu s$. This tail can be eliminated using a higher bias voltage.
No trace of saturation is evident from HPD7 and HPD8: in this case the charge
generated is much smaller.\\
A detailed study of the waveforms could be useful in the future to monitor
the beam quality in parasitic mode.\\
In fig. \ref{fig:fullstop_adc56} the histograms of the ADC counts (divided by
the attenuation factors) of HPD5 (left, solid line), of HPD6 (left, dashed
line) and their correlation plot are reported.
\par
The mean number of ADC counts (corrected for the attenuation factors) is
$(1.87\pm 0.03)\,10^{6}$ ADC counts for HPD5 and $(0.078\pm 0.001)\,10^{6}$
ADC counts for HPD6.\\
A calibration with the SM of the FRONT system gave the following results:
$(2.02 \pm 0.11)$ ADC counts/$\bar{p}$ for HPD5 and $(1.60 \pm 0.05)$ ADC
counts/$\bar{p}$ for HPD6 (the number of ADC counts is normalised to the
solid angle). A cross check with the previous calibration, by rescaling the
solid angles and the HPD gain, is consistent within $4.5\%$.

%%%%

\section{Future plans}
In the 2001 run all the 6 scintillators shown in fig. \ref{fig:beam_line}
will be used: the FRONT ones to detect possible beam losses, especially
during the steering of the extraction line, the BARREL ones to monitor the
beam intensity parasitically and the ENDCAP ones to measure the trapping
efficiency when the variable degrader will be ready.\\
Once the ejection line will be equipped with the new SP, a more precise
calibration will be done with the SM.\\
A detailed study of the HPD's waveforms will be useful in further
understanding the characteristics of the beam and to improve its quality, a
crucial item for the ATHENA physics.

%%%%

\section{Conclusions}
The beam detector of the ATHENA experiment has been described: it consists of
6 plastic scintillators, each coupled from one side to a proximity focused
HPD without preamplifier. This system, after a calibration, has been used to
monitor parasitically a bunched beam of $\simeq 1.2\times 10^{7}\,\bar{p}$s
with a width of $\simeq$ 250 ns (FWHM), to detect possible losses in the beam
pipe and to measure the $\bar{p}$ trapping efficiency. This detector is
linear and stable up to charge responses of $\simeq 100$ nC and it works with
no loss of efficiency in the stray field of the ATHENA solenoid. It is a
reliable device, giving the opportunity to vary its gain from zero to a few
thousands without any change neither in detection efficiency, nor in
linearity.\\
It is the first time that a HPD is used without preamplifier, taking
advantage of the huge amount of light, to exploit its huge dynamic range.

%%%%

\section{Acknowledgements}
We wish to thank prof. P. Benetti of the University of Pavia and V. Filippini
of the INFN of Pavia for their interesting discussions during the project
phase, S. Bricola and C. Marciano of the INFN of Pavia who built all the
mechanics, prof. A. Rotondi of the University of Pavia and D. Grassi of the
INFN of Pavia who helped us in the setting up of the detector and in its
characterisation.

%%%%%%%%%%%%%

%%%%%%%%%%%%%


\begin{thebibliography}{99}
%%%%  introduction  %%%%
\bibitem{ATHE1} M.H. Holzscheiter et al., {\em Antihydrogen Production
and Precision Experiments\/}, {\bf CERN/SPSLC 96-47, SPSLC/P302},
October 20, 1996.
%
\bibitem{Mike_Charl} M. Charlton et al.; {\em Phys. Rep.\/} {\bf 241}
(1994) 65.
%
\bibitem{GRAV} R.J. Hughes, M.H. Holzscheiter; Journal of Modern Optics
{\bf 39} (1992) 263.
%
\bibitem{AD1} S. Baiard et al., {\em Design Study of the Antiproton
Decelerator: AD\/}, {\bf CERN/PS 96-43 (AR)}, November 1996.
%
\bibitem{ECOOL2} M.H. Holzscheiter et al.; {\em Phys. Lett.\/} {\bf A214}
(1996) 279.
%
\bibitem{PENN2} M.H. Holzscheiter et al.; {\em Phys. Lett.\/} {\bf A129}
(1998) 38.
%
\bibitem{XP2020} PHILIPS data sheets.
%
%%%%  HPD  %%%%
\bibitem{HPD1} R. DeSalvo, {\em CLNS\/}
{\bf 82-92} Cornell University, Ithaca, 1987.
%
\bibitem{HPD2} L.K. van Geest et al., {\em Nucl. Instr. and Meth. A\/}
{\bf 310} (1991) 261-266.
%
\bibitem{HPD3} R. DeSalvo et al., {\em Nucl. Instr. and Meth. A\/}
{\bf 315} (1992) 375-384.
%
\bibitem{HPD4} H. Arnaudon et al., {\em Nucl. Instr. and Meth. A\/}
{\bf 342} (1994) 558-570.
%
\bibitem{HPD5} G. Anzivino et al., {\em Nucl. Instr. and Meth. A\/}
{\bf 365} (1995) 76-82.
%
\bibitem{HPD6} V. Filippini et al., {\em Nucl. Instr. and Meth. A\/}
{\bf 424} (1999) 343-351.
%
\bibitem{LHCb} E. Albrecht et al., {\em Nucl. Instr. and Meth. A\/}
{\bf 411} (1998) 249-264.
%
\bibitem{CMS} P. Cushman et al., {\em Nucl. Instr. and Meth. A\/}
{\bf 387} (1997) 107-112.
%
\bibitem{DEP} Delft Electronische Producten, Dwazziewegen 2, NL-9300
AB Roden, The Netherlands.
%
%%%%   Schottky Pickup  %%%%
%
\bibitem{SP1} C. Gonzales et al., {\em An ultra low-noise AC beam transformer
for deceleration and diagnostics of low intensity beams\/}, {\bf PAC '99, NY},
1999.
%
\bibitem{SP2} M.E. Angoletta et al., {\em The new digital receiver based
system for antiproton beam diagnostics\/}, {\bf CERN/PS 2001-044 (BD)}.
%
%%%%   beam calibration   %%%%
\bibitem{Activ} P. Lubinski et al., {\em Phys. Rev. Lett.\/}
{\bf 73}, 3199 (1994).
%
\end{thebibliography}
\end{document}